\documentclass[doublecol]{epl2} 
\usepackage[dvipsnames]{xcolor}
\usepackage{comment}

\title{Slip length for a viscous flow over spiky surfaces}
\shorttitle{Slip length for a viscous flow over spiky surfaces} 
\author{Alexei T. Skvortsov\inst{1} \footnote{Email: alexei.skvortsov@defence.gov.au} \and Denis S. Grebenkov\inst{2} \footnote{Email:denis.grebenkov@polytechnique.edu} \and Leon Chan \inst{3} \footnote{Email:lzhchan@unimelb.edu.au} \and  Andrew Ooi \inst{3}\footnote{Email:a.ooi@unimelb.edu.au}  }
\shortauthor{A. T. Skvortsov \etal}

\institute{                    
  \inst{1} Defence Science and Technology Group - Melbourne, VIC 3207, Australia\\
  \inst{2} Laboratoire de Physique de la Mati\`ere Condens\'ee,\\ CNRS – Ecole Polytechnique, Institut Polytechnique de Paris, 91120 Palaiseau, France\\
  \inst{3} Department of Mechanical Engineering, The University of Melbourne - Parkville; VIC 3010, Australia\\
}

 {Low-Reynolds-number (creeping) flows}

\abstract{
For a model of a 3D coating composed of a bi-periodic system of parallel riblets with gaps we analytically derive an approximate formula for the effective slip length (an offset from the flat surface at which the flow velocity would extrapolate to zero) as a function of the geometry of the system (riblet period,  riblet height, and relative gap size). This formula is valid for an arbitrary  fraction of gaps    (i.e from  narrow riblets to  narrow gaps) and agrees with the known analytical results for the 2D periodic coating of riblets without gaps. We validate our analytical results with the numerical solution of the equations of the viscous (creeping) flow over the riblets with gaps.
}

\begin{document}

\maketitle

\section{Introduction}

The viscous flow over surfaces covered by sharp elements (riblets, grooves,  spikes,  or pillars) has been the key component in many problems of microfluidics (lab-on-a-chip \cite{Rothstein_2010,Stone_2004}), geophysics (canopy flows \cite{Monti_2022}), and biomechanics (the so-called shark skin phenomenon \cite{Bechert_1989,Luchini_1991,Dean_2010,Martin_2014}). A spiky coating has a  remarkable (and, perhaps, counterintuitive) property of drag 
(shear stress) reduction of the viscous flow compared to a flat surface, although in the former case, the contact area between the fluid and solid is much higher \cite{Deyn_2022,Ran_2021,Lee_2016}. This property has made spiky coatings an attractive candidate for many practical applications (e.g., drag reduction of ships and drones \cite{Domel_2018}, 
 improvement of propeller performance \cite{Chen_2023}, micro-pump design \cite{Rothstein_2010,Huang_2016})  and stimulated many experimental and theoretical studies  \cite{Crowdy_2011a,Prosperetti_2007,Prosperetti_2007a,Ling_2017,Bazant_2008,Golovin_2017,Quere_2008,Crowdy_2011,Truesde_2006,Philip_1972,Lauga_2003,Asmolov_2012}. This is  an active area of research with extensive literature, see \cite{Deyn_2022,Ran_2021,Zeng_2022,Li_2023,Crowdy_2022,Modesti_2021,Crowdy_2022a,Yariv_2023},  and references therein.

The effect of a coating of complex morphology on  viscous flow has been conventionally quantified by a parameter called effective slip length \cite{Rothstein_2010,Bechert_1989,Crowdy_2011a,Bazant_2008,Philip_1972,Lauga_2003,Asmolov_2012}. This parameter can be introduced with the following arguments. Near a flat surface the velocity of flow is directed along the surface, it is zero at the surface (no-slip boundary condition)    and can be modeled by a linear profile
\begin{equation}
\label{EE0}
 v (y)  = J y,
\end{equation}
where $y$ is the distance from the  surface, constant $J$  is related to the friction drag at the surface $\tau = \mu dv/dz = \mu J$, and $\mu$ is fluid viscosity. With a coating of complex morphology, the flow just above and inside the coating can be very complex.  Nevertheless,  far from the surface the linear relation $v(y)$ is restored  but with an additional parameter $\lambda$    
\begin{equation}
\label{EE1}
 v (y)  = J (y - \lambda),
\end{equation}
where  $\lambda$ (can have either sign)  has a dimension of length and is called the effective slip length \cite{Rothstein_2010,Bechert_1989,Crowdy_2011a,Bazant_2008,Philip_1972,Lauga_2003,Asmolov_2012}.  This parameter is the aggregated measure of the effect of coating morphology on the hydrodynamic properties of the surface.  Condition $y =  \lambda$  corresponds to the fictitious coordinate at which the flow velocity would extrapolate to zero (relative to the surface $y =0$). Likewise, the parameter $\lambda$ can be introduced by postulating a radiation boundary condition at the surface \cite{Prosperetti_2007a}:
\begin{equation}
\label{EE1a}
 v   +  \lambda \frac{\partial v }{\partial y} =0.
\end{equation}

\begin{figure}
    \centering
    \includegraphics[width=50mm]{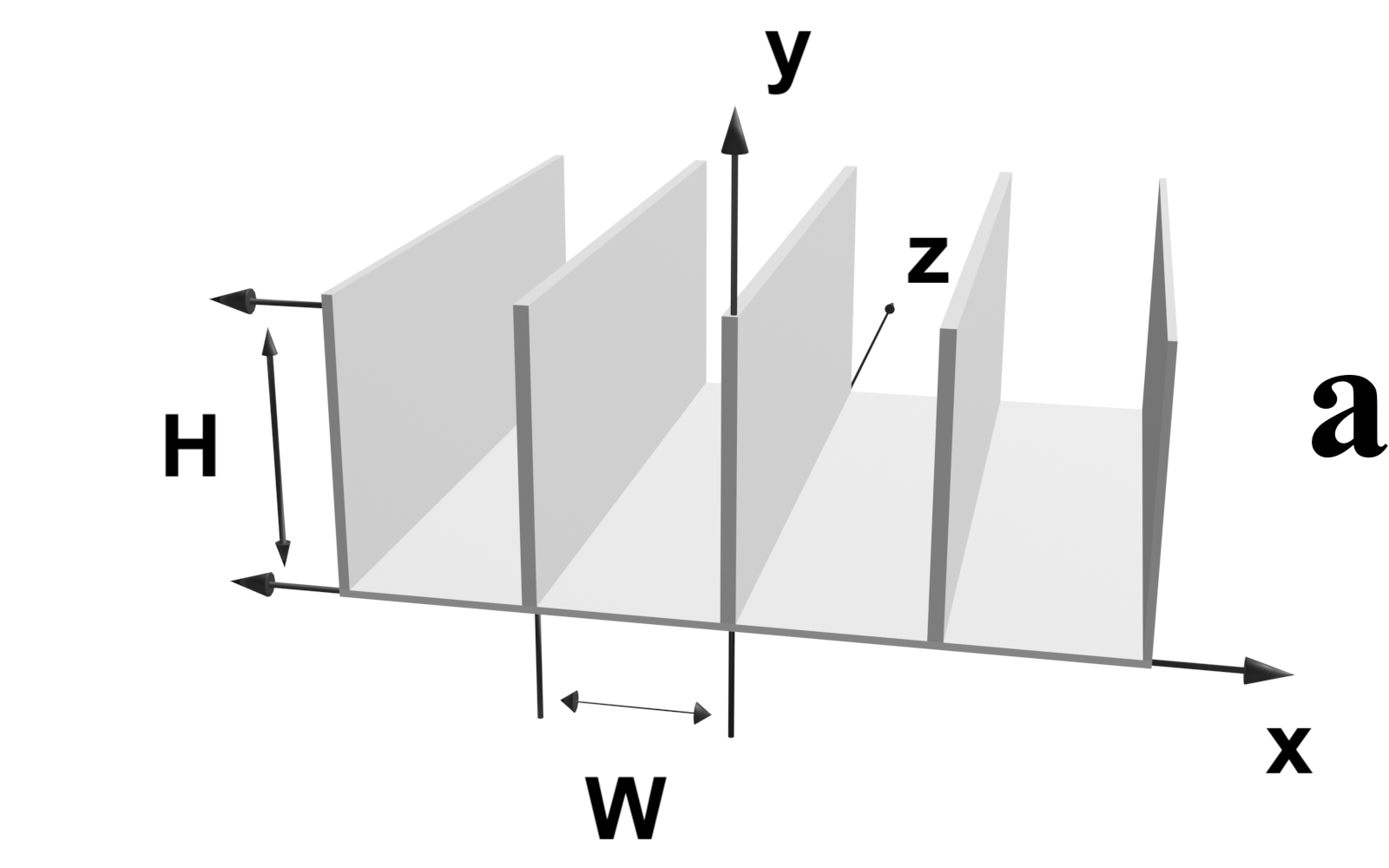}
    \includegraphics[width=50mm]{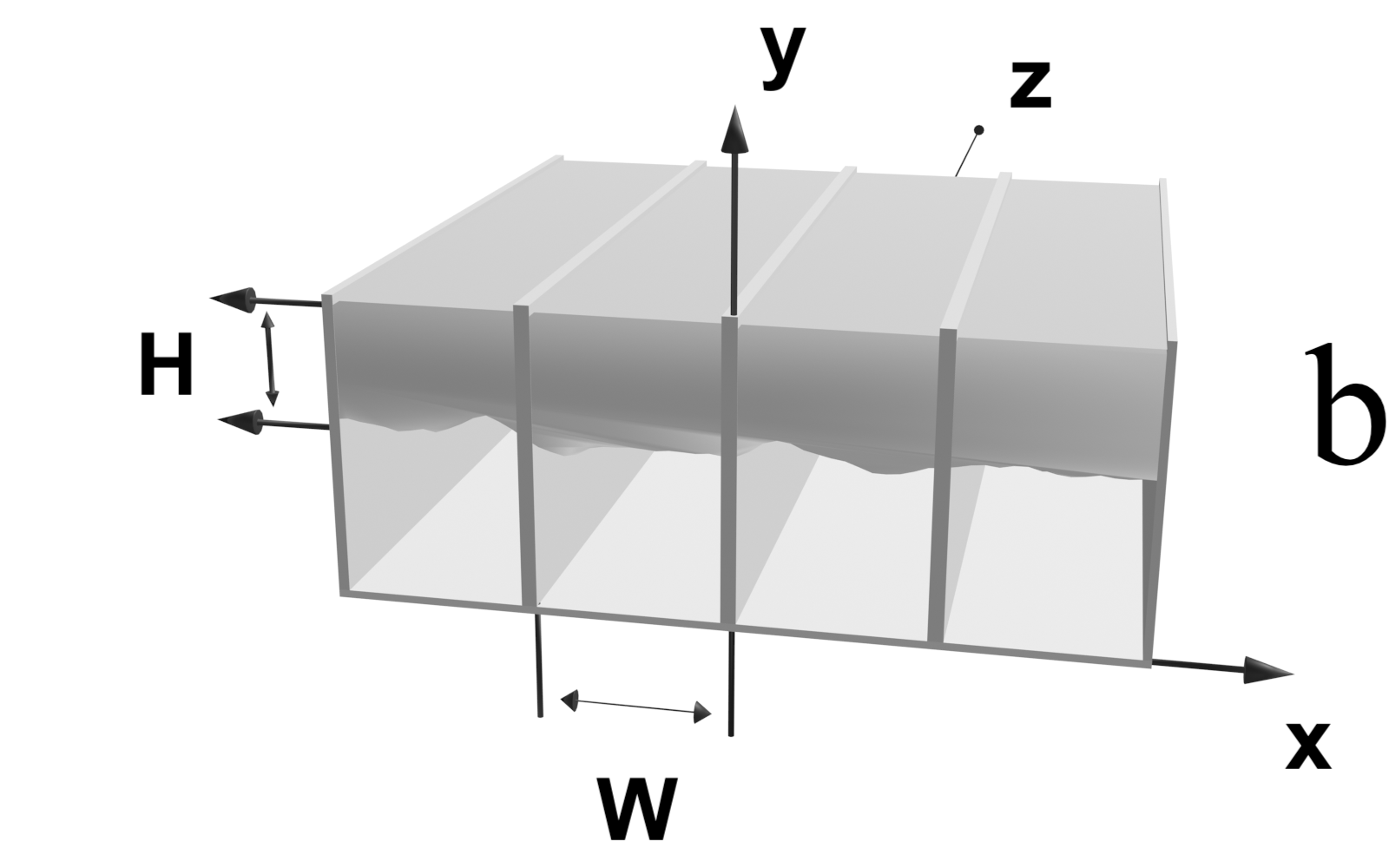}
    \includegraphics[width=50mm]{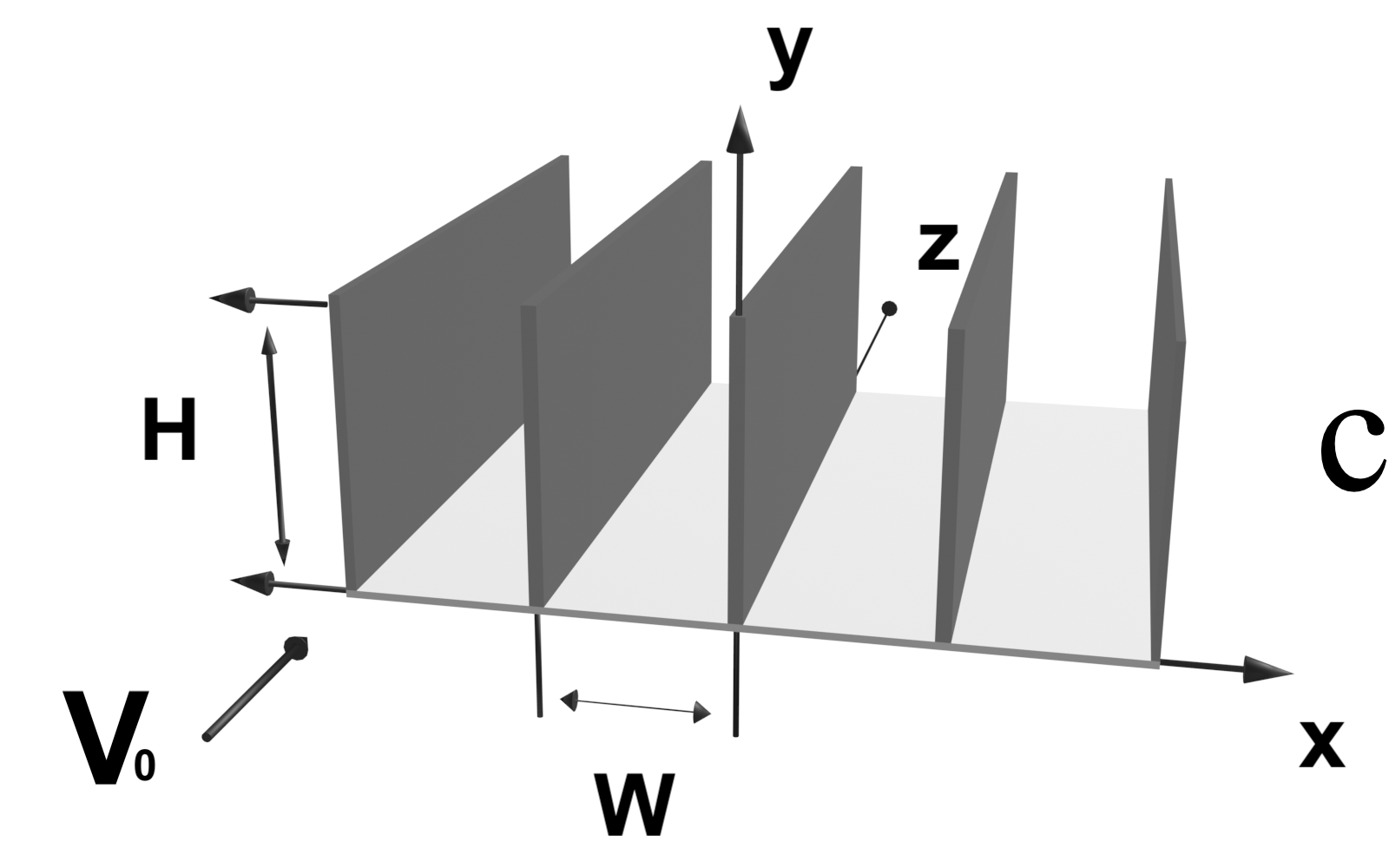}
    \caption{Examples of 2D riblet coating: \textit{a} - coating with a periodic system of riblets; \textit{b} - a  system of riblets with trapped air pockets; \textit{c} - coating with a periodic system of partially hydrophobic riblets modeled with partially absorbing (radiation) boundary condition (depicted in black); $\textbf{V}_0$ is the direction of the flow.}
\label{fig:Fig1}
\end{figure}

The effective slip length may also incorporate the effect of changing boundary conditions at some parts of the surface (from no-slip condition $v =0$ to no-stress condition $\tau =0$) due to the air bubbles trapped between the spikes, see Fig. \ref{fig:Fig1}. Evidently, the patches of no-stress areas of the surface (e.g., due to trapped air) may lead to a significant reduction of viscous drag and that is often referred to as hydrophobic properties of the coating. Alternatively, roughness can block access of the flow to some parts of the surface so that the intrinsic hydrophobicity of the surface (as its physicochemical property) can be significantly amplified \cite{Quere_2008}. This necessitates investigation of the interplay of the effects of hydrophobicity and roughness. To incorporate the effect of hydrophobicity of the spikes due to the no-stress and no-slip patches of the spike surface the spikes can be modeled with the radiation boundary conditions (see below).  All these cases are depicted in Fig. \ref{fig:Fig1}. 


The main focus of many theoretical studies of spiky coatings was the analytical derivation of the value of parameter $\lambda$ as a function of coating morphology. { Let us consider}  a steady  flow of viscous fluid with the low Reynolds number (Stokes flow). 
    The  flow is unidirectional and this implies that velocity vector is directed along the $z$ axis and depends only on the other two coordinates (i.e., $v  \equiv  v(x, y)$). It is well-known that in this case the equations  of motion reduce to the 2D Laplace equation  for the longitudinal velocity  \cite{Bechert_1989,Bazant_2016,Crowdy_2011a}
\begin{equation}
\label{E0}
\frac{\partial^2 v}{\partial x^2} + \frac{\partial^2 v }{\partial y^2}   = 0.
\end{equation}

For the 2D coatings (the canopy that is periodic in the cross-flow direction and does not change along the flow, see Fig.~1)  a wealth of analytical results have been derived by employing the property of conformal invariance of Eq. (\ref{E0}) \cite{Bechert_1989,Crowdy_2022,Luchini_1991}. For instance, for the 2D comb-like boundary or riblets shown on Fig. \ref{fig:Fig1}, the results are as follows. For the no-slip boundary condition at the spikes and on the base, Fig. \ref{fig:Fig1}a,   \cite{Bechert_1989,Skvortsov_2014}
\begin{equation}
\label{EE2}
\lambda  = \frac{W}{ \pi} \ln \left[{\cosh (\pi H/W)} \right]  \quad (\textrm{no-slip on the base}),   
\end{equation} 
and for  the no-slip boundary condition at the spikes and the no-stress (`hydrophobic')  boundary condition on the base, Fig. \ref{fig:Fig1}b,   \cite{Skvortsov_2018,Crowdy_2022}
\begin{equation}
\label{EE3}
\lambda = \frac{W}{ \pi} \ln \left[{\sinh (\pi H/W)} \right]  \quad (\textrm{no-stress on the base}), 
\end{equation}
where $W$ is the period of the comb-like structure (distance between spikes) and $H$ is the height of the spikes. As $H \rightarrow \infty$, both equations (\ref{EE2}) and (\ref{EE3}) behave similarly as
\begin{equation}
\label{EE4}
\lambda \approx H  - \frac{\ln 2}{\pi} W ,  
\end{equation}
so $\lambda$ tends to $H$ minus a universal offset proportional to the period of the structure.  For the riblets of other shapes  (e.g., semicircle, triangular, rectangular cross-sections) the results are similar and  can be found in  \cite{Bechert_1989,Luchini_1991}.

For a periodic configuration of alternating  (no-slip and no-stress) stripes on a flat surface oriented perpendicular to the flow velocity  \cite{Philip_1972,Lauga_2003,Asmolov_2012}, one has
\begin{equation}
\label{EE5}
\lambda = \frac{W }{ 2 \pi } \ln \left[{1/\sin \left(\frac {\pi}{2} \sigma \right) } \right], 
\end{equation}
where $\sigma$ is the surface fraction of the no-stress stripes, $W$ is the period of the stripes.

For the 3D morphological structures of the coating (e.g., spikes, pillars, or hemispheres) conformal transformation cannot be applied and there are only a limited number of papers in which the parameter $\lambda$ has been derived analytically, see \cite{Davis_2009,Davis_2010,Schnitzer_2018,Ng_2011,Ybert_2007,Lindsay_2017,Berezhkovskii_2004} and references therein. In particular, the authors of Refs. \cite{Davis_2009,Davis_2010,Schnitzer_2018,Ng_2011}  considered the model (that we refer to as the disk model) in which a viscous flow exists only above  `nanoforest'  composed of a lattice of identical cylindrical pillars. In this model,  the effect of the coating on the viscous flow is reduced to the friction forces acting at the top disks  of each pillar (the flow satisfies the no-slip boundary condition on
the top of the circular pillar, $y = 0$) whilst the flow inside the `nanoforest' (e.g., between the pillars) is disregarded (at $y = 0$ the no-stress boundary condition was assumed everywhere except the top disks).  It was found that at the limit of a small areal density of the `nanoforest' pillars (or surface fraction of the top disks), $\sigma \ll 1$, the  effective slip length  obeys the scaling law \cite{Davis_2009,Yariv_2023}
\begin{equation}
\label{EE6}
\lambda =  H - \left(\frac{A}{\sqrt{\sigma} }  - B \right) W,  ~~~ \sigma \rightarrow 0,
\end{equation}
where $W$ is the period of the pillar lattice (for simplicity assumed to be the square lattice),  $A= (3/16)\sqrt{\pi}$, $B =  (3/2 \pi) \ln (1 + \sqrt{2} ) \approx 0.4208$ \cite{Davis_2010} (note that this original value of $B$ was later corrected in \cite{Schnitzer_2018} to $B \approx 0.4655$).   It was also found that this scaling law is geometry specific, viz., for the elongated (quasi-one-dimensional) cross-section of the structural elements (wall-like in the current context) it changes from the power-law (\ref{EE6}) to the logarithmic form \cite{Davis_2010,Yariv_2023}: 
\begin{equation}
\label{EE7}
\lambda = H - \frac{1}{3 \pi} \ln \left(\frac {4} { \pi \sigma} \right) W,  ~~~ \sigma \rightarrow 0.
\end{equation}
Formulas  (\ref{EE6}) and  (\ref{EE7}) are  valid for $\sigma \ll 1$  \cite{Davis_2009}, \cite{Davis_2010}. The  limit $\sigma \to 0$ (no disks) corresponds to $\lambda \to -\infty$ or no friction (drag) on the surface  so that the approximate relation (\ref{EE1}) does not make sense anymore.  The later condition follows from Eq. (\ref{EE1a}) when the first term, which is proportional to $1/\lambda$,  becomes insignificant. The second terms in Eqs. (\ref{EE6}) and (\ref{EE7}) become zero at  
$\sigma = (A/B)^2 \approx 0.18$ and $\sigma = 4/\pi \approx 1.27$, respectively,  instead of $\sigma = 1$ (uniform surface with the no-slip boundary condition), which is due to the inapplicability of Eqs. (\ref{EE6}) and  (\ref{EE7}) at high $\sigma$.


The aim of the present paper is to derive the self-consistent {approximate} expression for the effective slip length of the spiky coating as a function of the height of its 3D structural element, similar to Eqs. (\ref{EE2}), (\ref{EE3}) for 2D.

\section{Riblets with periodic gaps}

\begin{figure}
\centering
\includegraphics[width=0.4\textwidth,height =0.4\textwidth]{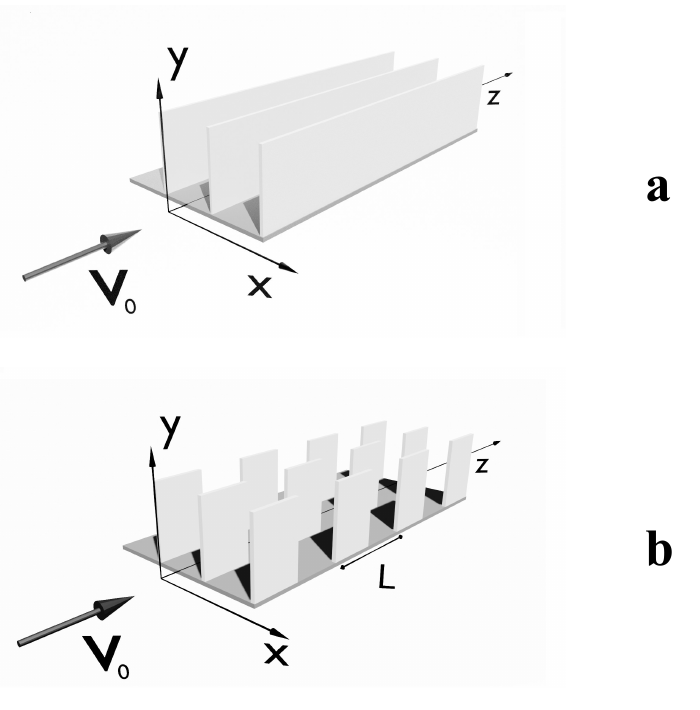}
\caption{Models of coatings: \textit{a} - periodic system of riblets; \textit{b} - periodic system of riblets with periodic gaps.}
\label{fig:Fig2}
\end{figure}

{ 

To appreciate the effect of the pillar height, we need to incorporate the flow between pillars. 
This flow indeed depends on the gaps between the pillars in the row. Without the gaps the solution is given by Eqs.~(\ref{EE2}, \ref{EE3}). To deduce an approximate model of the flow with gaps we use the well-known framework  of slip length that was conventionally applied for the analytical treatment of the  Stokes flow with the periodic boundary conditions \cite{Rothstein_2010,Philip_1972,Lauga_2003,Asmolov_2012}.   

Assume that the periodic system of pillars  that was formed by the 2D riblets with  the periodic identical gaps  as shown in Fig.~\ref{fig:Fig2}.
Near the edges of the gaps the flow has strong downstream dependency that rapidly (exponentially) disappears in the traverse direction, so that the flow between riblets becomes uniform in the downstream direction with its velocity being determined by the size of the gaps (for details, see \cite{Asmolov_2012}  and references therein). 

Let $x,y$ denote the horizontal and vertical axis, respectively, and the $z$ axis is directed along the riblets (and flow velocity) as shown in Fig. \ref{fig:Fig2}.   The boundary condition at the solid part of the riblets is $v =0$ (no-slip) and at the gaps, the boundary condition is $\partial_x v =0$ (no tangential stress ). Applying the aforementioned arguments, the alternating boundary conditions at the riblet surface ($y < H, x =  \pm W/2 $) imply that this surface (grey riblets in Fig. \ref{fig:Fig1}c)  can be translated to the problem of Stokes flow over a  texture of superhydrophobic transverse strips \cite{Rothstein_2010,Philip_1972,Lauga_2003,Asmolov_2012} and analytically  treated with the effective boundary condition
\begin{equation}
\label{E1}
 \frac{\partial v}{\partial x} + \frac{v}{\lambda_s}  = 0, ~~~ x = \pm W/2,~~~ 0 < y < H,
\end{equation}
where $\lambda_s$ is the slip length and given by Eq. (\ref{E2})}
\begin{equation}
\label{E2}
\lambda_s  = \frac{L }{ \pi } \ln \left[1/{\sin \left(\frac {\pi}{2} \sigma_s \right) } \right], 
\end{equation}
where $L = s + g$ is the period of the solid-gap structure of an individual riblet, $s$ is the width of the solid part of the riblet (per period), $g = L - s$ is the width of the gap,  $\sigma_s = s/L$. For the case $g =0$ (no gaps) we return to the solution given by Eqs. (\ref{EE2}, \ref{EE3}).   

With the effective 
boundary condition (\ref{E1}) the original 3D problem reduces to a 2D problem that can be tackled analytically (although due to the radiation boundary condition at the riblets conformal mapping is not helpful). Assume that $v = v_0 =const$ for some $y = \delta \gg H$ (Couette flow) and for a given $\lambda_s$ we can derive offset $\lambda$ as $\delta \rightarrow \infty$.
The parameter $\lambda$, being a function only of the geometry of the coating, is independent of $\delta, v_0$, and $\mu$.

Formally, due to the periodicity of the system, we need to find a solution of the Eq. (\ref{E0}) for $0 \le y \le \delta$ and $|x| < W/2$ with the following boundary conditions (see Fig. \ref{fig:Fig1}c):
\begin{equation}
\label{E3}
 \frac{\partial v}{\partial x} + \frac{ v}{\lambda_s}  = 0, ~~~ x = \pm W/2,~~~ 0 < y < H,
\end{equation}
\begin{equation}
\label{E4}
 \frac{\partial v}{\partial x}  = 0, ~~~ x = \pm W/2,~~~ H < y < \delta,
\end{equation}
\begin{equation}
\label{E5}
 v  = v_0, ~~~  y = \delta,
\end{equation}
\begin{equation}
\label{E6}
 v  = 0, ~~~  y = 0, ~~~\textrm{no-slip base,}
\end{equation}
or 
\begin{equation}
\label{E7}
\frac{\partial v}{\partial y} = 0, ~~~  y = 0, ~~~\textrm{no-stress base}.
\end{equation}

The parameter $\lambda$ for this setting can be derived from the solution of Eqs. (\ref{E3} -- \ref{E7}) by assuming that far above the coating  ($H \ll y \ll \delta$) the solution takes the form  (\ref{EE1}) and then by matching this solution with the one inside the coating ($0 \le y \le H$). {This is possible but would involve some tedious calculations \cite{Grebenkov22}.}
For the purpose of this study, we derive a  simpler (approximate) solution for $\lambda$ that can straightforwardly be deduced from the fact that the Robin boundary conditions  (\ref{E3}) can be replaced with the homogeneous boundary conditions $v=0$ but imposed on the equivalent boundaries at $y = W/2 + \lambda_s$ and $y = - W/2 - \lambda_s$. From here the approximate solution is immediately given by Eqs. (\ref{EE2}, \ref{EE3}) with substitution  $W \rightarrow W + 2\lambda_s$: 
\begin{equation}
\label{E8}
\lambda  = \frac{W + 2\lambda_s}{ \pi} \ln \left[{\cosh (\pi H/(W +2\lambda_s))} \right]   
\end{equation}
for the no-slip boundary condition on the base, and
\begin{equation}
\label{E9}
\lambda  = \frac{W + 2\lambda_s}{ \pi} \ln \left[{\sinh (\pi H/(W + 2 \lambda_s))} \right]. 
\end{equation}
for the no-stress on the base, where $\lambda_s$ is given by Eq. (\ref{E2}). 

\begin{figure}
    \centering
    \includegraphics[width=77mm]{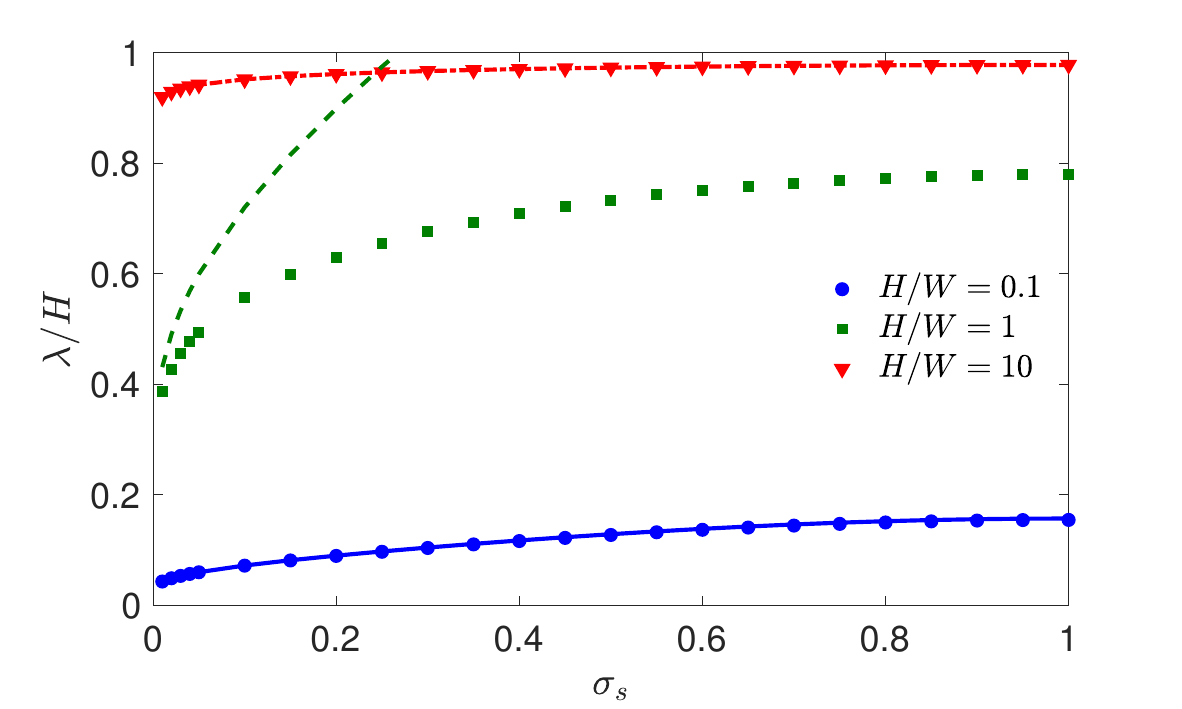}
    \includegraphics[width=77mm]{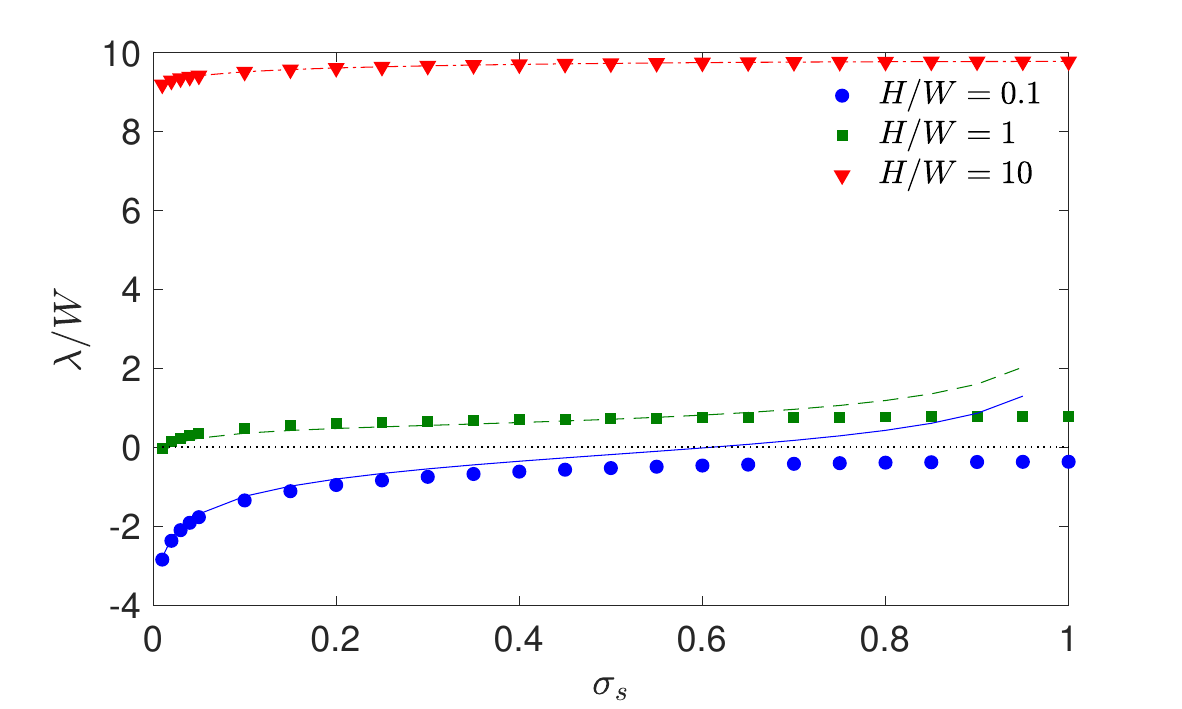}
    \caption{The effective slip length $\lambda$ as a function of $\sigma_s$, shown by symbols and given by Eqs. (\ref{E8}, \ref{E9}) for the no-slip (top) and no-stress (bottom) condition on the base, respectively.  Here $L/W = 1$ and three values of $H/W$ are considered.  Solid and dashed lines present the asymptotic behavior (\ref{E11}, \ref{E12}) for moderate heights $H/W = 0.1$ and $H/W = 1$, while dash-dotted line indicates the large-height asymptotic relation (\ref{E10}).  Note that $\lambda$ is rescaled by $H$ on the left panel and by $W$ on the right panel.}
    \label{fig:lambda}
\end{figure}

This is the main result of the present Letter. It provides insights into the dependence of the effective slip length on two-dimensional arrangements of the pillars and their height.   At $\lambda_s =0$ we return to the previous results for the 2D case,  Eqs. (\ref{EE2}), (\ref{EE3}). As $H \rightarrow \infty$, we recover the asymptotic relation similar to  Eq. (\ref{EE4}):  
\begin{equation}
\label{E10}
\lambda \approx H  - \frac{\ln 2}{\pi} (W+2\lambda_s).
\end{equation}
In view of Eq. (\ref{E2}) the limit $\sigma_s\to 0$ in this formula recovers the logarithmic dependency similar to Eq. (\ref{EE7}). Finally, for  $\lambda_s \rightarrow \infty$ and $H$ fixed (sparse configuration of the needle-like pillars), one finds
\begin{equation}
\label{E11}
\lambda  \approx  \frac{\pi H^2}{2(W+2\lambda_s)} ,   ~~~\textrm{no-slip  base}, 
\end{equation}
forno-slip  base, and
\begin{equation}
\label{E12}
\lambda \approx \frac{2 \lambda_s + W}{ \pi} \ln \left(\frac{ \pi H}{W+ 2 \lambda_s} \right) - \frac{W}{\pi}.
\end{equation}
for no-stress  base.
For the case of a square configuration of pillars ($W=L$) the plots of Eqs. (\ref{E8}, \ref{E9}) are depicted in Fig. \ref{fig:lambda}.
For the no-slip boundary condition on the base (top panel), the parameter $\lambda$ is positive and exhibits a monotonic increase with $\sigma_s$ (i.e., the fraction of the solid part of the riblet). 
When $H$ is large, the dependence on $\sigma_s$ is very weak, while $\lambda$ remains close to $H$, except for very small values of $\sigma_s$ (note that $\lambda$ is rescaled by $H$ in this panel). In particular, one sees that the asymptotic relation (\ref{E10}) accurately reproduces the full solution. In turn, in the limit $\sigma_s\to 0$ (no riblet), the effective slip length vanishes according to Eq. (\ref{E11}), as it should. However, this limit is achieved extremely slowly for large $H$.  In fact, one has $\lambda_s \approx (L/\pi) \ln (1/\sigma_s)$, and this parameter should be much larger than $\pi H/2$ or, equivalently, $\sigma_s \ll e^{-\pi^2 H/(2L)}$, to be able to apply Eq. (\ref{E11}).
For instance, if $H/L = 10$ (the red curve), one has $e^{-\pi^2 H/(2L)} \sim 10^{-22}$, i.e., the asymptotic relation (\ref{E11}) is not applicable for any reasonable $\sigma_s$.  In contrast, if the riblet height $H$ is much smaller than $W$, the asymptotic relation (\ref{E11}) provides an accurate approximation for the whole range of $\sigma_s$. In the intermediate case when $H \sim W$, the relation (\ref{E11}) is applicable only for small $\sigma_s$ (top panel, middle curve).

The situation is quite different for the no-stress boundary condition on the base (bottom panel). While the effective slip length still grows monotonously with $\sigma_s$, it takes negative values as $\sigma_s \to 0$. Moreover, $\lambda$ diverges
to $-\infty$ in this limit, in agreement with Eq. (\ref{E12}). As previously, the approach to this limit is very slow when $H$ is large, so that Eq. (\ref{E12}) is not applicable. In this setting, one can use the large-height expression (\ref{E10}).

The parameter $\lambda$ completely defines the relative change of drag due to coating 
\begin{equation}
\label{E13}
\frac{\tau_{\rm coat} -\tau_{\rm flat}}{\tau_{\rm flat}} =  \frac{\lambda}{\delta - \lambda} \approx \frac{\lambda}{\delta} 
\end{equation}
for $\lambda \ll \delta$ and $\delta$ is defined in Eq. (\ref{E5}). Eqs (\ref{E8}) and (\ref{E9}) and the data presented in Fig. \ref{fig:lambda} provide a clear guidance for targeted coating optimisation.

{
\section{Numerical Validation}

To validate our results we solve  { the 3D Stokes system of equations} of  fluid motion numerically by imposing the boundary condition of $v= v_0 = consts$ far away from the coating. Without the riblets the solution corresponds to the conventional  Couette flow (viscous flow with the linear velocity profile with no-slip boundary condition at $z=0$). The presence of  riblets changed the velocity profile in accordance with Eq. (\ref{EE1}) and this allows us to retrive the parameter $\lambda$. The solution was implemented with   the open source code NEK5000, a high-order spectral element solver \cite{nek5000_2008}. The Reynolds number of the flow based on the prescribed velocity at the top boundary of the flow ($v_0$) and the riblet height ($H$) was fixed at $Re_H= v_0 H/\nu = 10$ ($\nu$ is the kinematic viscosity of the medium), so that the Reynolds number based on the spanwise spacing of the riblets $Re_W = v_0 W/\nu$   ranges from 5 to 40. The number of hexahedral spectral elements in the simulations ranged from 90 to 120 elements and a fifth order polynomial with a Gauss-Lobatto-Legendre grid spacing within the element was used to fully resolve the flow. Selected cases were also simulated at a lower Reynolds number ($Re_H = 1$) (and the same riblets height) and with a refined mesh.  In all cases the same results were obtained. This  indicates that at this flow regime  (Stokes flow) and for simulated scenarios  the numerical results are independent of  Reynolds number and mesh resolution. The value of the effective slip length $\lambda$ was determined by extrapolating the linear velocity profile from far above the riblets to the position corresponding   to  $v= 0$. The effective slip length is simply  the distance between the tip of the riblets and this position.

Top panel of Fig.  \ref{fig:Numerics}
compares numerical  and analytical results in Eq. (\ref{E8}). The bottom panel presents the relative deviation (in percentage) between the analytical and numerical results.  The best agreement is observed for the case of low riblets   (the left part of the figure), 
which is almost independent of gaps in the riblets’ walls. In turn, the worst agreement corresponds to high riblets with larger gaps in their walls (right bottom corner). This trend is intuitively clear, since our simplified assumption of the vertical wall homogenisation  implies a  uniform velocity of the flow which obviously cannot be satisfied for the very high riblets (flow localises near the tips of the spiky coating  without penetrating to its bottom).

\begin{figure}
    \centering
    \includegraphics[width=77mm]{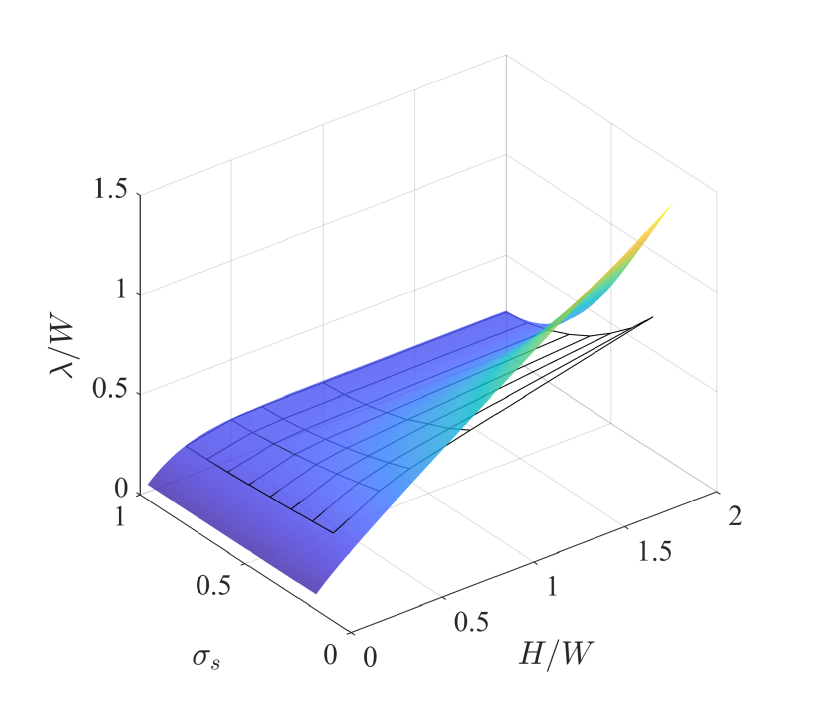}
    \includegraphics[width=77mm]{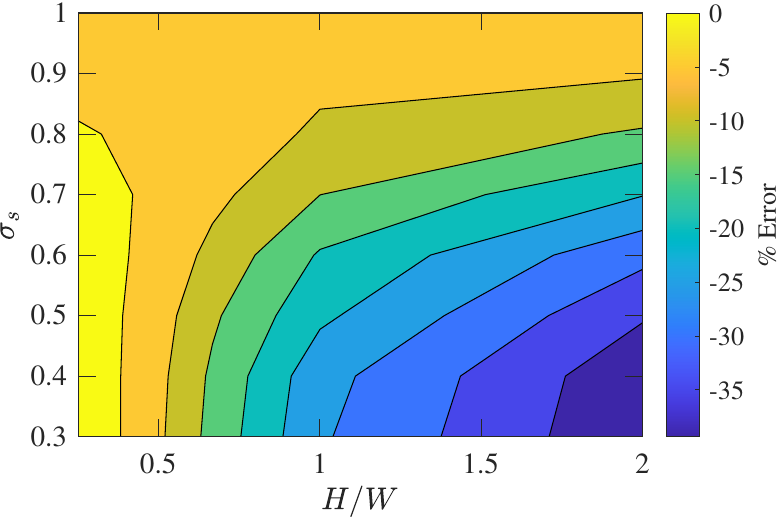}
    \caption{Comparison of analytical prediction  for the effective slip length given by Eq (\ref{E8}) with the result derived from the numerical solution of the equations of  viscous  flow over coating  with riblets.  The top panel  depicts  the plots of solutions and the bottom figure is their relative deviation (percentages).}
    \label{fig:Numerics}
\end{figure}

}

\section{Discussion and future work}

In summary, for a model of 3D spiky coating  we derived an approximate formula for the effective slip length as a function of the pillar height, and the 2D arrangements of the pillars.
{
For the  case of unidirectional flow over the riblets without gaps   the parameter $\lambda$ is a scalar. With the presence of the gaps in the riblets's wall the flow becomes three-dimensional and  parameter $\lambda$ becomes a tensor. Due to apparent symmetry arguments, this tensor reduces  to the two components, viz., along  and across the flow. 
Our approach assumed that the two-dimensional structure of the flow 
is approximately preserved, so the cross-flow component of $\lambda$ is relatively small and can be disregarded.
}

There are several extensions that can easily be incorporated into the proposed model. For instance, there is no need to assume that riblets should have only one gap per period, since there is the formula for $\lambda_s$ for an arbitrary number of gaps per period  \cite{Crowdy_2011,Skvortsov_2020}. This enables the analytical treatment of coatings with much more complex structures.

Our results can also be extended for the pillars of arbitrary cross-section (i.e., different from an infinitely thin interval)  provided the momentum flux through the top surface of the pillar  can still  be neglected. To this end, we can apply the following rationale. It is known that the Stokes force (which is proportional to the momentum flux over the surface of the pillar) is proportional to the capacitance of the object (or logcapacity in 2D) \cite{Hubbard_1993}.  As a consequence, to translate the results of the proposed framework to arbitrary pillars, it is sufficient to find a solid interval of an equivalent logcapacity for a given pillar cross-section (e.g., the logcapacity of an ellipse with semi-axes $a$ and $b$ is $(a + b)/2$ \cite{Landkof_1972}). Moreover, as the parameter $s$ is in the argument of logarithm, the final result is insensitive to minor inaccuracies in estimation of the equivalent logcapacity. Indeed, this approximation can only hold for sparse configurations: $s/W \ll 1$.

This approach also allows us to make informative conclusions regarding the applicability of the disk model\cite{Davis_2009,Yariv_2023} for a coating of tall pillars (a brush). 
In the disk model all drag is generated by the viscous flow acting on top of the pillars while in the present model
it is a result of viscous force acting on their side surface  of the riblets whilst the contribution of the force from the top surface area  is neglected.
The effective slip length due to the momentum flux through the top surface and the side of the pillars can be characterised by the second terms (offsets)  in Eqs. (\ref{EE6}), (\ref{E10}), respectively. By comparing these terms we arrive at the simple condition of validity of the disk model
\begin{equation}  
\label{E15}
\frac{1}{\sqrt{\sigma}}  \ll \frac{(1 + 2\lambda_s/W) \ln 2}{\pi A}   + \frac{B}{A},
\end{equation}
where constants $A$ and $B$ are defined in Eq. (\ref{EE6}) and $\lambda_s$ is given by Eq. (\ref{E2}).  We note that both terms on the right side of this inequality are positive and   $B/A \approx 5.4$ so this condition is quite restrictive for $\sigma$.

We believe that the presented results can be useful for the targeted design of engineered coatings with desirable hydrodynamic properties before proceeding with extensive computational simulations and experimental evaluation.

\section{Data availability}

The data that support the findings of this study are available from the corresponding author upon request.

\section{Acknowledgements}

A.T.S. grateful to Ian R. MacGillivray and Paul A. Martin and  many insightful discussions.  D.S.G. acknowledges the Alexander von Humboldt Foundation for support within a Bessel Prize award.

\end{document}